# The pulse shape of cosmic-ray ground-level enhancements

**H. Moraal[1]**

*Centre for Space Research, School for Physical and Chemical Sciences, North-West University, Potchefstroom, 2520, South Africa.*

*E-mail: harm.moraal@nwu.ac.za*

**K.G. McCracken**

*Institute for Physical Science and Technology, University of Maryland, College Park, MD, 20742, U.S.A*

*E-mail: jellore@hinet.net.au*

**R.A. Caballero-Lopez**

*Ciencias Espaciales, Instituto de Geofísica, Universidad Nacional Autónoma de México, 04510 México D.F., México.*

*E-mail: rogelioc@gmail.com*

Enhancements of the comic-ray intensity as observed by detectors on the ground have been observed 71 times since 1942. They are due to solar energetic particles accelerated in the regions of solar flares deep in the corona, or in the shock front of coronal mass ejections (CMEs) in the solar wind. The latter is the favoured model for the classical "gradual" ground-level enhancement (GLE). In several papers since the one of McCracken et al. (2008), we pointed out, however, that some GLEs are too impulsive to be accelerated in the CME shocks. With this hypothesis in mind we study the time profiles of all the available GLEs. The main results are that there is a continuous range from gradual to impulsive, that the fastest risers are concentrated at heliolongitudes that are magnetically well-connected to Earth, and that the shape of the pulse is a powerful indicator of propagation conditions between Sun and Earth. This ranges from relatively quiet to highly disturbed.

---

[1]Speaker

## 1. Introduction

It is well-known that ground-level enhancements (GLEs) in the intensity of cosmic rays as measured by neutron monitors are associated with solar flares and coronal mass ejections (CMEs), and that they originate primarily from western longitudes on the surface of the sun.

Studies of the time structure of solar energetic particle events as measured in space, generally classify them as "impulsive" or "gradual", e.g. Reames (1999). Flares in the low solar corona are relatively short-lived and have dimensions much smaller than one solar radius. On the other hand, CMEs develop more gradually at distances beyond about four solar radii, are much larger than the sun, and should therefore have shock fronts that are widely extended in heliolatitude and longitude. It is therefore natural to associate the impulsive events with acceleration in solar flares, and gradual events in CME shock fronts.

The earth's neutron monitor network is sensitive to the arrival direction of the particles, which gives an indication of the anisotropy of the event. For vertical arrival at a neutron monitor, the particles must have come from a so-called asymptotic direction in space before they penetrated the geomagnetosphere. From the knowledge of the asymptotic cones of acceptance of the world's ~ 40 neutron monitors, this network provides a means to view the events in two dimensions, and infer the direction of propagation of the anisotropic beam of particles. This directional sensitivity explains why a given GLE can display an impulsive character on one set of neutron monitors, but only show up as gradual on others, as observed by, e.g. Shea and Smart (1990) and Miroshnichenko et al. (2000).

McCracken et al. (2008) used this directional sensitivity of the network on the very large GLE 69 on 20 January 2005. It was such that three neutron monitors (South Pole, Terre Adelie and McMurdo) saw an extremely impulsive increase, reaching the peak intensity in ~ 5 minutes, while Thule saw a much longer rise time of ~ 30 minutes. A third group of stations observed two consecutive peaks. This is shown in Figure 1 of our accompanying paper Moraal et al. (2015a). McCracken et al. interpreted this structure as indicating two injection/acceleration mechanisms, namely a fast, short-lived solar flare, and a slower, longer-lived bow shock of the CME. Moraal and McCracken (2012) then applied this analysis to the 16 GLEs of solar cycle 23 and identified six more such double-peak events. McCracken et al. (2012) subsequently pursued all the impulsive GLEs over the entire time span, and demonstrated that these events typically extend to higher energies than the more gradual ones.

Finally, Moraal and Caballero-Lopez (2014) analysed the large GLE 42 on 29 September 1989 in such a way that it puts the concept of impulsive vs. gradual GLEs in perspective. It was shown that GLE 42 was much less anisotropic than GLE 69. The solar activity associated with GLE 69 was in the "well-connected" region at ~ 65$^\circ$ W, while for GLE 42 it was invisible behind the western limb and could only be inferred indirectly at ~ 120$^\circ$ W. This longitude would have been so poorly connected that particles that might have been accelerated in the highly anisotropic first beam in the lower corona, could not have reached Earth.

In this paper we study the pulse shapes or time profiles of all GLEs simultaneously, to determine how to differentiate between impulsive and gradual ones, and how to interpret its physical meaning. In order to do this, we start in Figure 1 to show the well-known pattern of the number of GLEs as function of heliolongitude of the inferred source of the event. The distribution peaks in the heliolongitude range of 30º to 75º W. This distribution has been interpreted since the papers of McCracken et al. (1962 a,b,c). The observations are from the data base described by Moraal et al. (2012), which contains all available observations of all 71 GLEs observed since 1942. Only GLEs with amplitude > 10% are used here. The distribution in Figure 1 is similar to that of Smart and Shea (1996), for the events that were observed until then. These authors, however, also included GLEs < 10%. Gopalswamy et al. (2012) extended this study to 16 GLEs that were observed in solar cycle 23, from GLE number 55 on 6 November 1997 onwards. The same trend was observed for this subset, with the mean and median longitudes 44º W and 59º W. However, the statistical significance was weaker.

In this study we record the maximum increase, the time-to-maximum, the time to decay to 50% from the maximum intensity, and the heliolongitude of the inferred source of each event. We then interpret these observations in terms of a simple diffusive model. The results indicate that there is a continuous range of pulses ranging from impulsive to gradual; that the most impulsive ones originate from the heliolatitude range 30º W to 60º W; that the relationship between rise and decay time of the pulses broadly agrees with the model for diffusive propagation; and that estimates of the cosmic-ray diffusion mean-free-path between the sun and earth can be inferred from this.

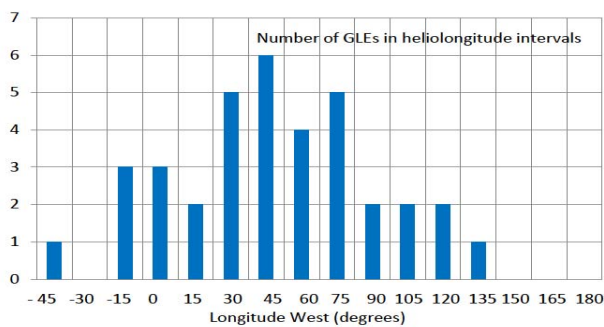

**Figure 1.** The number of GLEs with amplitude > 10%, observed in 15º intervals of solar longitude.

## 2. Observation of the pulse shape of GLEs

Figure 2 displays the rise time of the events with amplitude >10% as function of heliolongitude.

For each of the events the rise time of only one neutron monitor was used, selected as follows: First, the observing station has to be at a geomagnetic cutoff rigidity $P_c$ < 1 GV, so that the magnitude of the event is determined by the atmospheric, and not geomagnetic cutoff. This eliminates energy/rigidity dependence in the selection. Second, the intensity increase is usually anisotropic, so that the magnitude of the event depends on the asymptotic direction (i.e. after correction for magnetospheric bending) of viewing of the detector. In all cases the station with the largest increase was chosen. To read off the rise times, the beginning of the event was taken as the earliest time for which a statistically significant increase could be identified by eye from the intensity-time profile. The time of maximum was generally easy to read off, but was sometimes complicated when the profile lingered around the maximum intensity, or when it even show more than one maximum.

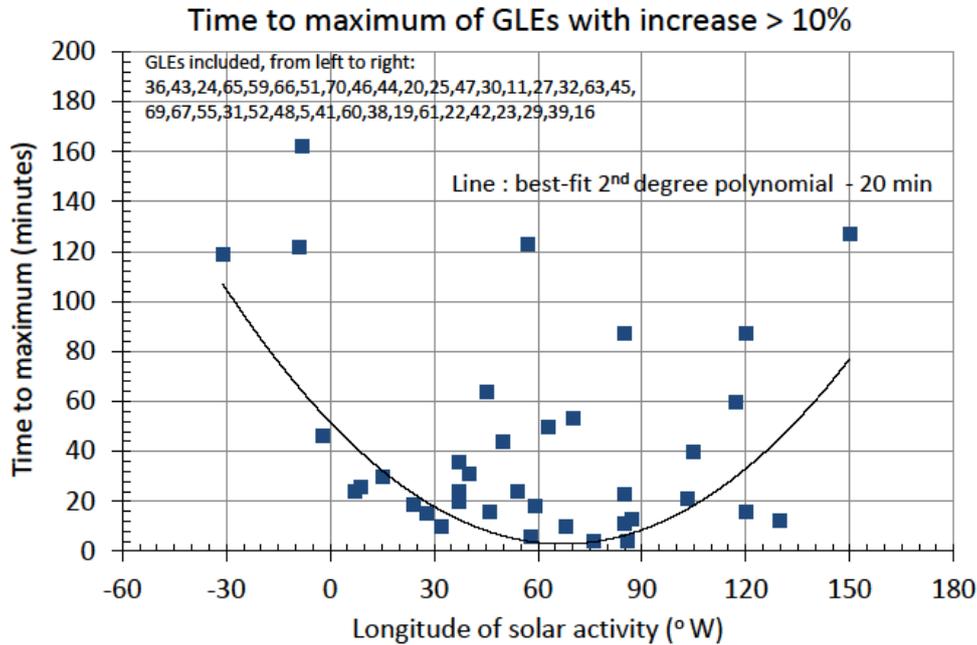

**Figure 2.** Time to maximum intensity of 37 GLEs with amplitude > 10%. The sequence of GLEs from left to right is noted on the figure. The line is the best-fit second-order polynomial, minus 20 minutes.

There is a large scatter in the distribution of Figure 2, but there is also a clear tendency that the most impulsive events are concentrated in the "magnetically well-connected" heliolongitude region. A best-fit second-order polynomial was fitted to the data, and the minimum increase time occurs at ~ 65° W. The curve shown on the graph is this polynomial, but with 20 minutes subtracted. This curve is guides the eye through the minimum rise times. The main interpretation is that the most impulsive events occur in the magnetically well-connected region because it is particles from flare acceleration that are observed. For other latitudes this impulsive peak is less visible at Earth, and it is only the particles accelerated in the CMEs that are seen.

GLEs 29 and 39, the third and second last from the western end, are notable exceptions. Their rise time is significantly below the expected minimum. These events had amplitudes of 10 and 100% respectively, and close inspection showed nothing peculiar in their time profiles. They are therefore interpreted as two events in which the Parker field was significantly distorted, so that the field from far western longitudes was well-connected to Earth.

In addition to the rise time, the decrease profile for each event was also recorded. This was defined as the time interval from maximum intensity to 50% of that value. It is possible to choose later times, such as the e-folding time or even deeper, but then the procedure becomes increasingly difficult due to larger statistical uncertainties, and also due to variations caused by temporal changes in propagation conditions.

Figure 3 shows the relationship between the rise and decay times (to 50% of the peak intensity)

of all the GLEs in Figure 2. The 10 largest ones are shown in red. They are predominantly impulsive, but GLEs 42 and 52 are noteworthy exceptions. The full line on the plot is the best-fit linear regression line. This shows that on average the decay time is about twice as long as the rise time.

The main result of this analysis is that GLEs do not separate into two distinct categories of impulsive and gradual. The rise times show a continuous range from 5 to 160 minutes.

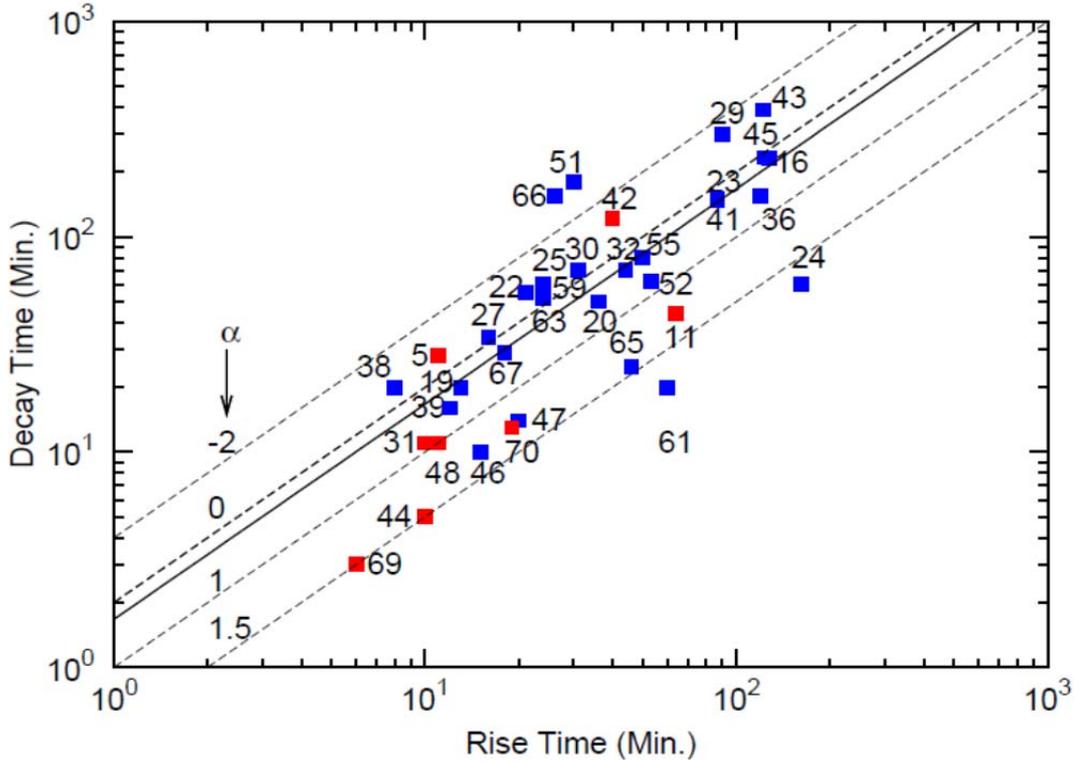

**Figure 3.** The relationship between rise time and decay time (to 50% of the peak intensity) for GLEs > 10%. The full line is the best-fit linear regression line, while the four dashed lines show the expected relationship for a point-diffusion model as explained in equation (5) below.

In the next section we interpret these results in terms of a diffusive transport model.

## 3. Interpretation in terms of a diffusive transport model

In Moraal et al. (2015b) we argue that the convective and adiabatic loss terms in the Parker transport equation are small relative to the diffusive term. In this case the transport equation reduces to the standard diffusion equation

$$\frac{\partial f}{\partial t} = \frac{1}{r^2} \frac{\partial}{\partial r}\left[ r^2 \kappa \frac{\partial f}{\partial r} \right].$$
(1)

If the diffusion coefficient is of the form $\kappa = \kappa_0 (r/r_0)^\alpha$, the solution for an impulsive injection at $t = 0$ and $r = 0$ is (e.g. Duggal, 1979):

*The pulse shape of cosmic-ray ground-level enhancements*

$$f \propto t^{3/(\alpha-2)} \exp\left[\frac{-r^2(r_0/r)^\alpha}{(2-\alpha)^2 \kappa_0 t}\right] \quad (2)$$

We call this the point-diffusion solution.

The increasing phase of this time-intensity profile is determined by the second, exponential part, while the decay phase is determined by the first power-law part. The increasing phase depends on both the magnitude $\kappa_0$ and the radial dependence $\alpha$ of the diffusion coefficient. The decay phase, however, depends only on $\alpha$.

It is readily shown that this profile reaches its peak at the time

$$t_p = \frac{r^2(r_0/r)^\alpha}{3(2-\alpha)\kappa_0}, \quad (3)$$

and that the peak intensity is $f_p = (e\, t_p)^{3/(\alpha-2)}$. In terms of $t_p$ and $f_p$ the profile (2) can be written as

$$\frac{f}{f_p} = \left[\frac{t}{t_p}\exp\left(\frac{t_p}{t}-1\right)\right]^{\frac{3}{\alpha-2}}. \quad (4)$$

The time to reach half the peak intensity is the solution of the transcendent equation

$$\frac{t_{1/2}}{t_p} = 2^{\frac{2-\alpha}{3}} \exp\left(1 - \frac{t_p}{t_{1/2}}\right). \quad (5)$$

This has two solutions, for $t < t_p$ and $t > t_p$. They are the rise and decay times to half the maximum intensity. It is noteworthy that this ratio is independent of the radial distance, $r$, and the magnitude of the diffusion coefficient, $\kappa$. It only depends on the radial dependence of $\kappa$. For $\alpha = 0, 1,$ and $2$, corresponding to $\kappa \propto r^0$, $\kappa \propto r^1$ and $\kappa \propto r^2$, the solution of (5) is $t_{1/2}/t_p = 3.13; 2.15;$ and $1.07$ respectively. These values can conveniently be parameterized as $t_{1/2}/t_p \approx 3 - \alpha$.

This relationship is shown by the dashed lines in Figure 3. The vertical axis of this figure actually shows the decay time $t_d = t_{1/2} - t_p$, and with the above parameterization this is given by $t_d/t_p = 2 - \alpha$. Note that as $\alpha$ approaches the value 2, the ratio $t_d/t_p$ approaches zero. This corresponds to a free-escape scenario.

These lines are almost perfectly parallel to the observed regression average. The regression average corresponds to a diffusion coefficient of the form $\kappa \propto r^{0.25}$. This weak radial dependence is somewhat unexpected. It is widely observed that diffusion coefficients vary inversely proportional to magnetic field strength. In the inner heliosphere, for $r \ll 1$ AU, the Parker spiral magnetic field for quiet, undisturbed conditions is quasi-radial, and it falls off as $r^{-2}$. For $r \gg 1$ the fall-off of the field goes as $r^{-1}$, so that the fall-off at Earth ($r = 1$ AU) should be intermediate between these two values. Hence, for undisturbed heliospheric conditions one expects that the data points should lie in the region $1 < \alpha < 2$, but only 9 GLEs do so. This then, is interpreted as that for most of the GLEs, the heliosphere is in a disturbed state, where $\kappa$ increases much slower than proportional to $r^2$ or $r^1$.

Li and Zank (2005) pointed out that this is possible due to a series of CMEs that fill space between the sun and Earth. Li et al. (2012) explicitly discussed a model of twin (or multiple) CMEs. The first CME accelerates particles and leaves increased magnetic turbulence in its wake. The follow-on CME then expands in an environment where the upstream turbulence is high and the diffusion coefficient small. This causes efficient shock acceleration.

## 4. Discussion

The main results of the paper are that there is no clear distinction of GLEs in impulsive and gradual classes, but rather a continuous range between these extremes, and that the time profiles of GLEs can be interpreted in a simple point-diffusion model.

This interpretation of the rise and decay times is preliminary because our model is oversimplified. First, the process of convection and adiabatic energy should be included in the full transport equation, instead of the analytical solution of the simplified equation (1) used here. Both processes will have an effect on the magnitude of the pulse. However, they should not affect the shape of the pulse (i.e. the $t_d/t_p$ ratio in Figure 3), because both processes are non-dispersive in time, so that this should not distort the GLE shape.

A second limitation of the model is that the increases are typically not spherically symmetric, in particular in the beginning (rising) phase. Particles will leak out of the beam in the latitudinal and longitudinal directions, and this will change both the rise and decay times. Since this effect should be largest in the beginning stages of the event, will make the ratio $t_d/t_p$ decrease with time. This implies that points in Figure 3 should shift to the right. The amount of shift should be inversely proportional to the angular extent of the source region. This is equivalent to a larger $\alpha$, which relaxes the need for strong turbulence conditions ahead of the CME to produce the observed time profiles.

On the other hand, a prolonged injection will place the data points in Figure 3 further to the right relative to the lines for impulsive injection. This will decrease the effective value of $\alpha$.

These modifications and effects will be studied in subsequent papers, mainly by comparing the time profiles of individual events with observations of the heliospheric magnetic field and solar wind, and their disturbances during the events.

## Acknowledgements

This work was supported by South African NRF Grant SNA2011110300007, and Mexican PAPIIT-UNAM grant IN110413. K.G. McCracken acknowledges the consistent support he has received since 2005 from the International Space Science Institute (ISSI), Bern, Switzerland.The neutron monitor observations used are from the worldwide network, collected in the database described in McCracken et al. (2012) http://usuarios.geofisica.unam.mx/GLE_Data_Base/files/. This data base originated from the work of Shea et al. (1985, 1987), it was managed by Shea, Smart and colleagues until 1999, and updated and migrated by M.L. Duldig, (Duldig and Watts, 2001), E.A. Eroshenko and H. Moraal.

*The pulse shape of cosmic-ray ground-level enhancements*